\documentclass[twocolumn,showpacs,preprintnumbers,amsmath,amssymb,nofootinbib]{revtex4}
\usepackage{verbatim}
\usepackage{slashed}
\usepackage{epsfig}
\usepackage{graphicx}
\usepackage{amsmath}
\usepackage{mathrsfs}
\usepackage{bm}

\begin{document}


\title{$J/\psi$ Production Via Photon Fragmentation at Hadron Colliders}


\author{Zhi-Guo He, Rong Li and Jian-Xiong Wang}
\affiliation{Institute of High Energy Physics, Chinese Academy of
Science, P.O. Box 918(4), Beijing, 100049, China\\
Theoretical Physics Center for Science Facilities, CAS,
Beijing, 100049, China.}



\date{\today}

\begin{abstract}
The transverse momentum ($p_t$) distributions of production and polarization for $J/\psi$ measured by CDF Collaboration are still challenging our understanding of the heavy quarkonium production mechanism even with recent significant theoretical progresses on the next-to-leading-order (NLO) QCD calculation. We suggest a new mechanism for $J/\psi$ production at hadron collider, $pp(\bar p)\rightarrow \gamma^{\ast}(J/\psi)+X$ with $J/\psi$ from a virtual photon $\gamma^{\ast}$ fragmentation. Our calculations show it's $p_t$ distribution at NLO will be larger than that of the conventional $J/\psi$ production from the color-singlet mechanism at NLO when $p_{t}>26$ (35) GeV at the Tevatron (LHC) and reach about 6 (10) times of the conventional one when $p_t=50$ (100) GeV at the Tevatron (LHC), in spite of a suppression factor $(\alpha/\alpha_s)^2$ that is associated with the QED and QCD coupling constants. In addition, it also has large impact on the $p_{t}$ distribution of $J/\psi$ polarization in large $p_{t}$ region. Therefore, it is a important mechanism for $J/\psi$ production at large $p_t$ region especially for the LHC. 
\end{abstract}

\maketitle


As good probes for people to understand both the perturbative and
non-perturbative aspects of QCD, the production of heavy quarkonium
in high energy processes has attracted considerable attention in recent years(see\cite{Brambilla:2004wf} for reviews). 

Within the framework of an effective field theory non-relativistic QCD (NRQCD) \cite{Bodwin:1994jh}, 
the orders of magnitude discrepancies between experimental data measured by the CDF Collaboration\cite{Abe:1997jz} and the color-singlet mechanism (CSM) theoretical predictions were well resolved by including the color-octet effect\cite{Braaten:1994vv}. However, the NRQCD prediction\cite{Beneke:1995yb} of the dominate transverse polarization for $J/\psi$ at sufficient large transverse momentum ($p_{t}$) is in bad agreement with the almost un-polarized results measured by CDF collaboration \cite{Affolder:2000nn,Abulencia:2007us}. To reveal the $J/\psi$ production mechanism, recently much theoretical effort has been made and some substantial progress has been achieved. At Hadron collider, the next-to-leading order (NLO) QCD corrections for conventional $J/\psi$ production $gg\to J/\psi g$ in the CSM are calculated in Ref~\cite{Campbell:2007ws} and the results show the $p_{t}$ distribution is significantly enhanced in large $p_t$ range; $gg\to J/\psi c\bar{c}$ in the CSM was found to make sizable contribution to $p_t$ distribution~\cite{Artoisenet:2007xi}; It is reported in Ref~\cite{Gong:2008sn} that the polarization for the conventional $J/\psi$ production via the CSM turns from much transverse at leading-order (LO) into much longitudinal at NLO, while the $J/\psi$ produced via $^{1}S_{0}^{(8)}$ and $^{3}S_{1}^{(8)}$ color-octet states still dominates transverse polarization at large $p_{t}$ at NLO\cite{Gong:2008ft}; It is argued in Ref~\cite{Haberzettl:2007kj} that the s-channel cut contribution may be a possible solution to the $J/\psi$ production puzzle. At $e^+e^-$ collider, the NLO QCD corrections~\cite{Zhang:2006ay} to $e^+e^-\rightarrow J/\psi c\bar{c}$ and $e^+e^-\rightarrow J/\psi gg$ in the CSM can well resolve the large discrepancies between the experimental data and LO theoretical prediction. Some other attempts may be found in Ref.\cite{Brambilla:2004wf}. Even with these recent significant theoretical progresses, the challenge to our understanding of the heavy quarkonium production mechanism is still exist. 

Since $J/\psi(1^{--})$  can couple with a photon, $pp(\bar p)\rightarrow \gamma^{\ast}(J/\psi)+X$ with $J/\psi$ from a virtual photon $\gamma^{\ast}$ fragmentation can  contribute to $J/\psi$ hadroproduction. It is thought to be small at first sight, and has never been taken into consideration.  However, there are really a few examples in which the kinematic enhancement can compensate the suppression factor $(\alpha/\alpha_{s})^{2}$ that is associated with the QED and QCD coupling constants. For example, the cross section is very large in $J/\psi$ electromagnetic production at $e^{+} e^{-}$ colliders \cite{Chang:1997dw}. $J/\psi c\bar{c}$ produced in $e^{+}e^{-}$ annihilation through two virtual photons \cite{Liu:2003zr} will prevail over that through one virtual photon when the center-of-mass energy $\sqrt{s}>20$ GeV.  And the cross section for $e^{+}e^{-}\to J/\psi J/\psi$\cite{Bodwin:2002fk} at LO is also a few times larger than that of $e^{+}e^{-}\to J/\psi \eta_{c}$\cite{Braaten:2002fi} at $\sqrt{s}=10.6$ GeV, although its QCD corrections is negative\cite{Gong:2008ce}. 
It is easy to see that the $p_t$ distribution for the conventional $J/\psi$ production from the CSM at LO scales like $1/p_{t}^{8}$ and  that for the photon fragmentation $J/\psi$ production at LO is $1/p_{t}^{4}$, which may substantially enhance the contribution of the later at large $p_{t}$ region. Furthermore, the contribution of the QCD corrections for the conventional one $gg\to J/\psi g$ behaves as $1/p_{t}^{6}$, so the $(m_{c}/p_{t})^2$ kinematic enhancement of this photon fragmentation production might even compensate the suppression factor $\alpha^{2}/\alpha_{s}^{3}$ compared to the former when $p_{t}$ is large enough. Therefore in this letter, we study the photon fragmentation $J/\psi$ production and polarization up to NLO in $\alpha_{s}$ at the Tevatron and LHC. The Feynman Diagram Calculation package (FDC)\cite{Wang:2004du} is used in the calculation.

 According to NRQCD factorization approach\cite{Bodwin:1994jh}, $J/\psi$ production rate at hadron collider is expressed as:
\begin{eqnarray}
\sigma[pp\to J/\psi+X]=\sum_{i,j,n}\int dx_{1}dx_{2}G_{i/p}G_{j/p}
\nonumber\\ \times
\hat{\sigma}[i+j\to(c\bar{c})_{n}+X]\langle\mathcal{O}^{H}_{n}\rangle,
\end{eqnarray} where $p$ is either a proton or antiproton. The short distance part $\hat{\sigma}$ represents the partonic production of $c\bar{c}$ with quantum number n.  And $\langle\mathcal{O}^{H}_{n}\rangle$ is the non-perturbative long distance matrix elements that parametrize the transition of the $c\bar{c}$ pair into $J/\psi$. 

\begin{figure}
\center{
\includegraphics[scale=0.48]{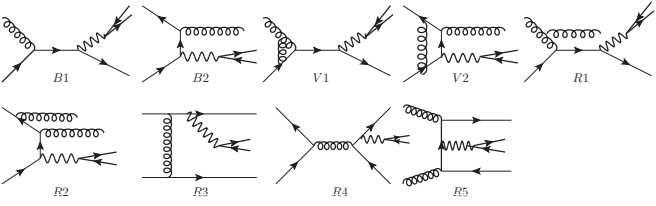}
\caption{Typical Feynman Diagrams for $J/\psi$ production through photon
fragmentation at LO(B1, B2), virtual corrections(V1,V2) and
real corrections(R1--R5).}
}
\end{figure}

At LO ($\alpha^{2}\alpha_{s}$), there are two partonic processes:
\begin{eqnarray}
B1:~g(p_{1})+q(p_{2})\to J/\psi (p_{3})+q(p_{4}),\nonumber \\
B2:~q(p_{1})+\bar{q}(p_{2})\to J/\psi (p_{3})+g(p_{4}),\nonumber
\end{eqnarray} where q represents all possible light quark u, d, s and anti-quark $\bar{u},\bar{d},\bar{s}$. The typical Feynman diagrams are shown in Fig.1. The partonic tree level results of the two processes for certain light q are trivial and given as:
\begin{eqnarray}
\frac{\mathrm{d}\hat{\sigma}_1}{\mathrm{d}t}= \frac{2\pi^2 \alpha^2
e_c^2 e_{q}^2\alpha_s \langle\mathcal{O}_{1}^{J/\psi}\rangle} {3m_c^{5}s^2}
\frac{((1-s)^2+(1-u)^2)}{su},\nonumber\\
\frac{\mathrm{d}\hat{\sigma}_2}{\mathrm{d}t}= \frac{2\pi^2 \alpha^2
e_c^2 e_q^2\alpha_s \langle\mathcal{O}_{1}^{J/\psi}\rangle} {3m_c^{5}s^2}
\frac{((1-u)^2+(1-t)^2)}{ut}
\end{eqnarray}
with $s=\frac{(p_1+p_2)^2}{4m_c^{2}}$,
$t=\frac{(p_1-p_3)^2}{4m_c^{2}}$, $u=\frac{(p_1-p_4)^2}{4m_c^{2}}$, 
$m_c$ is the mass of charm quark.

  The NLO correction is of $\alpha^{2}\alpha_{s}^{2}$ order. It contains the virtual(V) part and the real(R) part. At this order, the $^3S_1^{(1)}$ state could couple to one photon or one photon and two gluons. Only the one photon diagrams have kinematic enhancement at large $p_{t}$ region, for which the representative ones are displayed in Fig.1. And they also form a gauge-invariant subset. At present, we drop the contribution of the diagrams with $c\bar{c}$ pair coupling to three gauge bosons. The complete NLO corrections and their interference with full QCD processes will be discussed in further work.

  There are 19 fragmentation diagrams among all 33 NLO virtual correction diagrams for each Born process including the counter term diagrams. And the partonic virtual corrections for the differential cross section are:
\begin{equation}
\frac{\mathrm{d}\hat{\sigma}^V}{\mathrm{d}t}\propto
2\mathrm{Re}(M^{B}M^{V\ast}).
\end{equation}
The ultraviolet (UV) and infrared (IR) divergences will appear in the calculation of $M^{V\ast}$. $D=4-2\epsilon$ dimensional regularization is applied in all the calculations. We adopt the same renormalization scheme as in Ref.\cite{Klasen:2004tz}. The Coulomb singularity is isolated by the small relative velocity $v$ and then is absorbed by the long-distance matrix element. The IR divergences will disappear when including the real corrections.

There are a few subprocesses in the real corrections. We divided
them into seven categories below:
\begin{equation}
\begin{array}{lll}
&gg\to J/\psi q\bar{q}, &\quad q\bar{q}\to J/\psi gg,\nonumber\\
&q\bar{q}\to J/\psi q\bar{q}, &\quad q\bar{q}\to J/\psi q^{\prime}\bar{q}^{\prime}, \nonumber\\
&qq^{\prime}\to J/\psi qq^{\prime},
  &\quad \bar{q}\bar{q}^{\prime}\to J/\psi \bar{q}\bar{q}^{\prime},\nonumber \\  
&gq(\bar{q})\to J/\psi gq(\bar{q}), &\nonumber
\end{array}
\end{equation}
where $q,q^{\prime},(\bar{q},\bar{q}^{\prime})$ denote light quark(anti-quark) with different flavors. For all these processes, there will be soft and collinear poles in the phase space integration. We separate them in $D$ dimensions for coherence using the two-cutoff phase space slicing method\cite{Harris:2001sx}, in which the phase space is partitioned into three parts, the soft region, hard collinear region and hard non-collinear region by two small parameters: $\delta_{s}$ the soft cutoff and $\delta_{c}\ll\delta_{s}$ the collinear cutoff. Then the real corrections turn to be:
\begin{equation}
\sigma^{R}=\sigma^{S}+\sigma^{HC}+\sigma^{HC}_{add}+\sigma^{H\bar{C}},
\end{equation}
\begin{figure*}
\begin{center}
\includegraphics[scale=0.28]{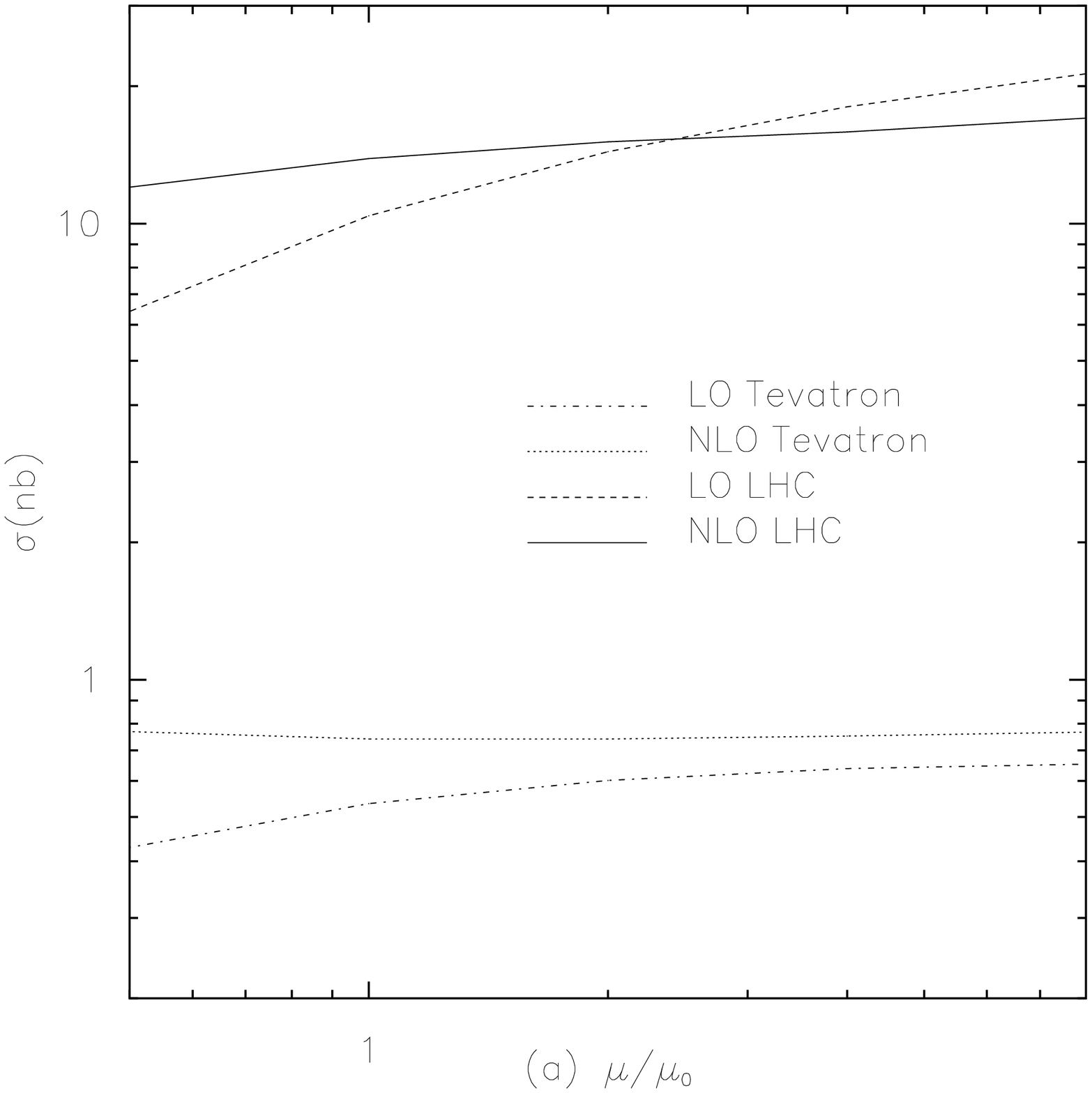}
\includegraphics[scale=0.28]{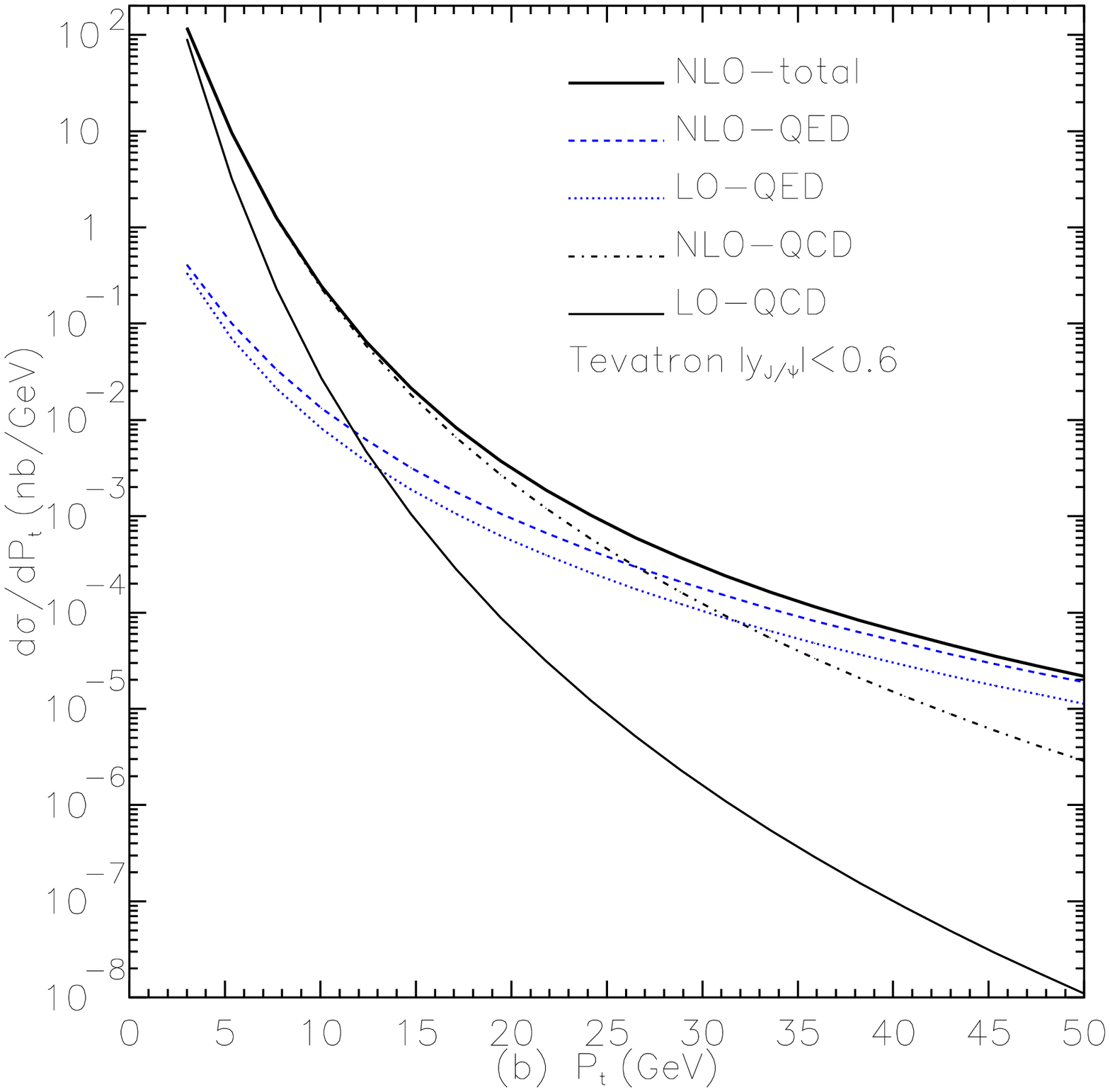}
\includegraphics[scale=0.28]{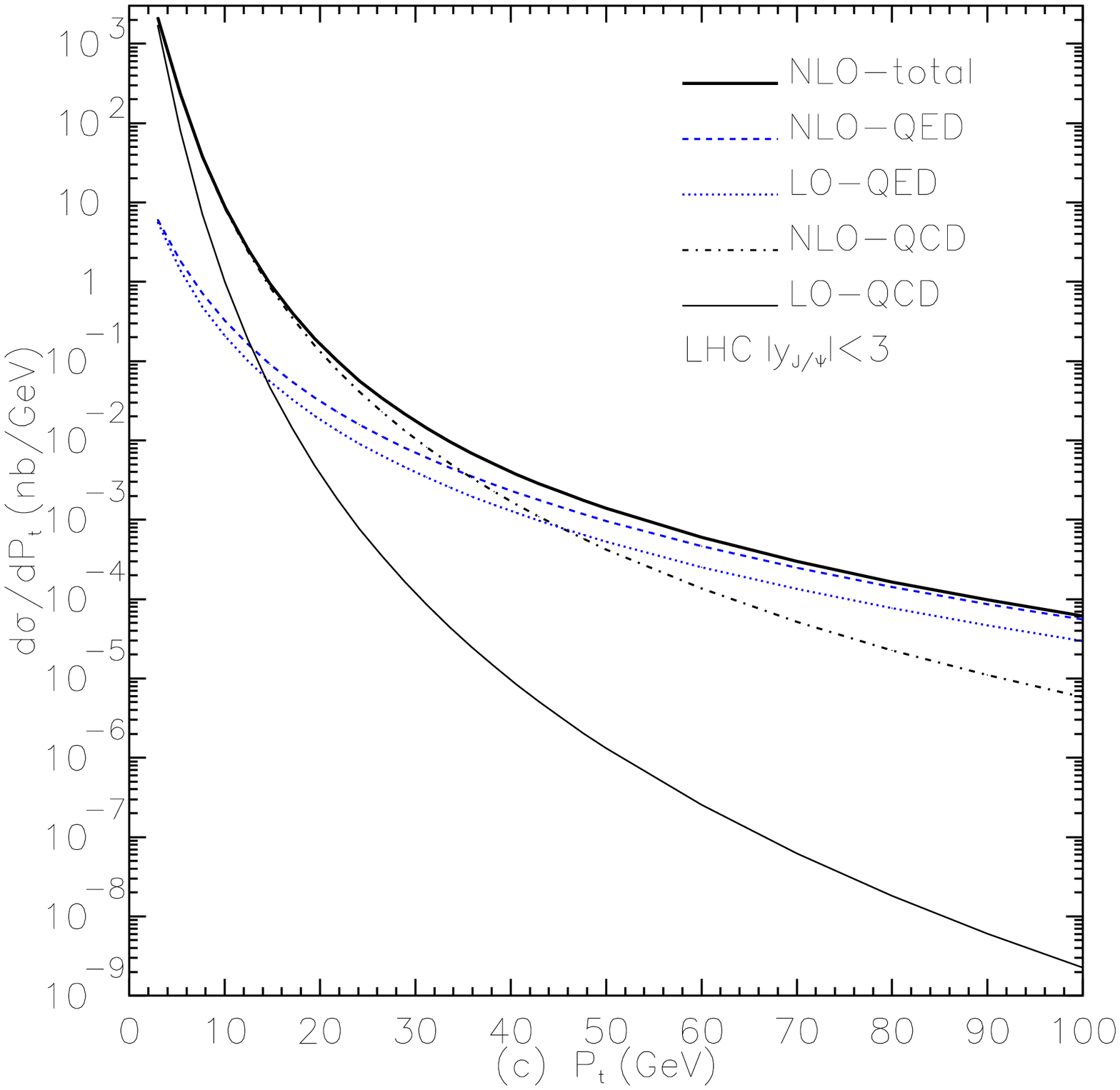}
\caption{(a)~The $\mu$ dependence of the cross section for the photon fragmentation $J/\psi$ production at the Tevatron(lower) and LHC(upper) at LO(dashed) and NLO(solid) with $\mu_{r}=\mu_{f}=\mu$. The $p_{t}$ distribution for $J/\psi$ production at the Tevatron(b) and LHC(c). ``QED" refers to the photon fragmentation production, ``QCD" refers to the conventional production from the CSM, and ``total" refers to the QED+QCD. The center-of-mass energies are 1.98 TeV at Tevatron and 14 TeV at LHC.}
\end{center}
\end{figure*} where $\hat{\sigma}^{S}$\footnote{Thereafter $\hat{\sigma}$ represents the corresponding partonic cross section.} is the contribution of soft region including soft divergences and is calculated analytically in the limit of the soft gluon. Note that the soft poles are all associated with light quarks or gluons, for only the one photon fragmentation diagrams are considered. $\hat{\sigma}^{HC}$ from the hard collinear region contains collinear singularities which are also factorized out in D-dimensions. The initial state collinear poles are absorbed into the redefinition of the parton distribution function (PDF)(traditionally called mass factorization\cite{Altarelli:1979ub}). Here a scale dependent PDFs with $\overline{MS}$ convention\cite{Harris:2001sx} are used. After the redefinition of PDF, there will be an additional part $\sigma^{HC}_{add}$ left. The final state collinear poles together with the soft ones will cancel that of the virtual corrections, $i.e.$ $\sigma^{S}+\sigma^{HC}+\sigma^{V}$ is finite. And the cancellation is also completed analytically. The last part $\sigma^{H\bar{C}}$ is finite and computed numerically.

To obtain the $p_{t}$ distribution of the $J/\psi$ production and polarization, the integral variables are transformed from $\mathrm{d}x_{2}\mathrm{d}t$ to $J \mathrm{d}p_{t}\mathrm{d}y$. Then we have:
\begin{equation}
\frac{\mathrm{d}\sigma}{\mathrm{d}p_{t}}=\int J
dx_{1}dyG_{\alpha}(x_{1},\mu_{f})
G_{\beta}(x_{2},\mu_{f})\frac{\mathrm{d}\hat{\sigma}}{\mathrm{d}t},
\end{equation} where y is the rapidity of $J/\psi$ in the laboratory frame and $\mu_{f}$ is the factorization scale. And the polarization parameter $\alpha$ is defined by:
\begin{equation}
\alpha(p_{t})=\frac{\mathrm{d}\sigma_{T}/\mathrm{d}p_{t}-2\mathrm{d}\sigma_{L}/\mathrm{d}p_{t}}
{\mathrm{d}\sigma_{T}/\mathrm{d}p_{t}+2\mathrm{d}\sigma_{L}/\mathrm{d}p_{t}}.
\end{equation}
To calculate $\alpha(p_{t})$, the $J/\psi$ polarization vectors $\epsilon(\lambda)$ are kept through our calculation which makes the calculation become more complicated. And the expression of the polarized $J/\psi$ partonic differential cross section could be explicitly written as:
\begin{equation}
\frac{\mathrm{d}\hat{\sigma}_{\lambda}}{\mathrm{d}t}=a
\epsilon(\lambda)\cdot\epsilon^{\ast}(\lambda)+
\sum_{i,j=1,2}a_{i,j}p_{i}\cdot\epsilon(\lambda)p_{j}\cdot\epsilon^{\ast}(\lambda),
\end{equation}
where $\lambda=T_1,T_2,L$. $\epsilon(T_{1})$,$\epsilon(T_{2})$ and $\epsilon(L)$ are the two transverse polarization vectors and the longitudinal one for $J/\psi$ respectively. Meanwhile, the sum over polarizations of the other particles is done in D-dimensions. One could find that $a$ and $a_{i,j}$ will be finite if the virtual and real corrections are properly dealt as mentioned before. To ensure the validity of our results, we also checked the gauge invariance by replacing the gluon polarization vectors with its momenta in the numerical calculation.

In the numerical calculations, the CTEQ6L1 and CTEQ6M PDFs\cite{Pumplin:2002vw} and the corresponding fitted values $\alpha_{s}(M_{Z})=0.130$ and $\alpha_{s}(M_{Z})=0.118$ are used for LO and NLO predictions respectively. QED coupling constant $\alpha=1/128$ and $m_{c}=M_{J/\psi}/2\simeq1.5$GeV are chosen. The long distance matrix element $\langle\mathcal{O}_{1}^{J/\psi}\rangle=1.35$GeV$^{3}$ is extracted from the leptonic decay of $J/\psi$ at QCD NLO level.  The renormalization scale $\mu_r$ and factorization scale $\mu_f$ are used as $\mu_r=\mu_f=\mu_0=\sqrt{(2m_c)^2+p_t^2}$. The two phase space cutoffs $\delta_s=10^{-3}$ and $\delta_c=\delta_s/50$ are chosen, and the invariance for different values of $\delta_s$ and $\delta_c$ is obviously observed within the error tolerance. All the results are restricted to the NRQCD applicable domain $p_t>3$ GeV, and $|y_{{J/\psi}}|<3$ for the LHC, $|y_{{J/\psi}}|<0.6$ for the Tevatron.

\begin{figure*}
\center{
\includegraphics[scale=0.30]{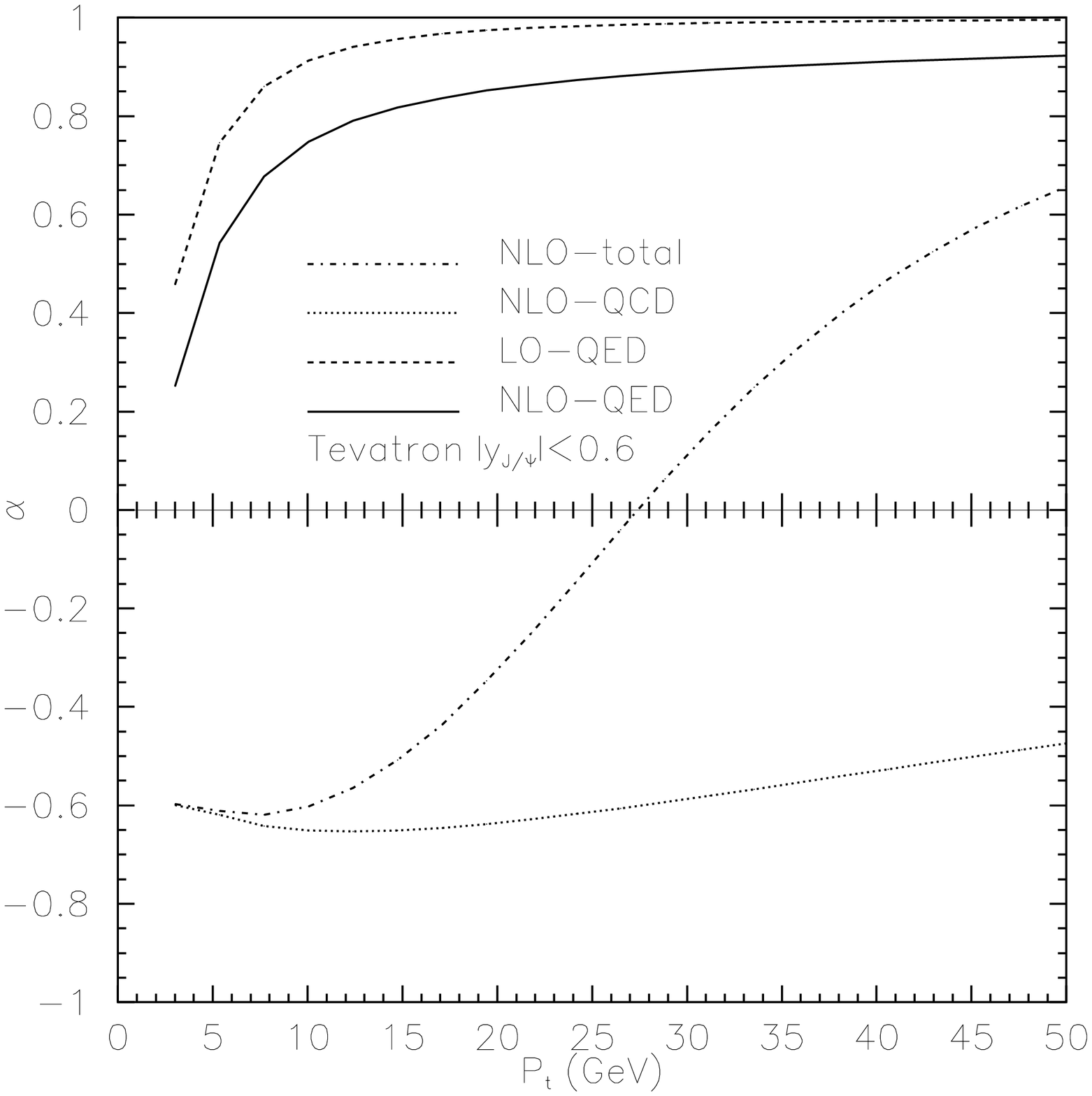}~~~~~~~~~~~~
\includegraphics[scale=0.30]{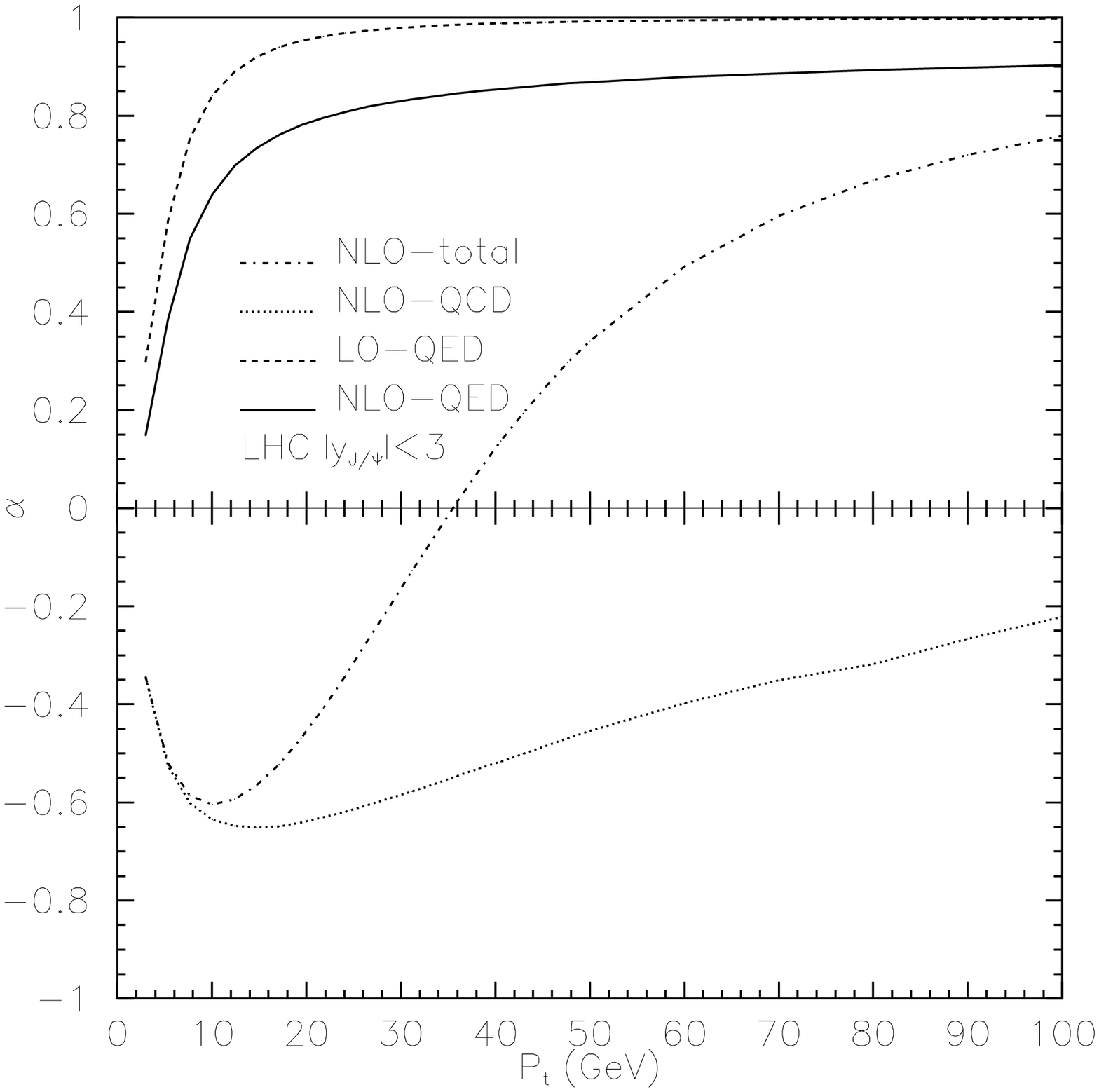}
\caption{The $p_{t}$ dependence of the polarization parameter $\alpha$ for $J/\psi$ production at Tevatron(left) and LHC (right).}
}
\end{figure*}

The numerical results for the photon fragmentation $J/\psi$ production are presented in Fig.~2 and 3. For comparison, the plots for the conventional $J/\psi$ production from the CSM at NLO are also shown in the two figures, and they will be referred as the conventional one in the following statement. The curves in Fig.~2.a show that the scale dependence is well improved at NLO at both the Tevatron and LHC, and the QCD corrections enhance the total cross section about $30\%\sim40\%$ for our default choice of scale $\mu_r=\mu_f=\mu_0$.  From Fig.~2.b, it is clear that the $p_{t}$ distribution of $J/\psi$ production for the new mechanism at NLO is comparable to the conventional one at middle $p_{t}$ range ($p_{t}\simeq20$GeV), and is larger than the conventional one as $p_{t}>26$GeV, and reaches about 6 times of the conventional one when $p_t=50$ GeV. The similar results for the LHC are shown in Fig.~2.c. The $p_t$ distribution for the new one at NLO is larger than the conventional one as $p_t>35$ GeV and reaches about 10 times of the conventional one at $p_t=100$ GeV. The $p_t$ distribution of $J/\psi$ polarization at the Tevatron and LHC are shown in Fig. 3.  As can be seen from both the Tevatron and LHC results, because of $J/\psi$ from photon fragmentation, the polarization parameter $\alpha$ at LO is positive and turn close to 1 quickly when $p_{t}$ increase.  And its QCD corrections only make $\alpha$ slightly lower down. It has been given in Ref.\cite{Gong:2008sn} that the $J/\psi$ polarization at NLO is mainly longitudinal for the conventional one. When the contributions for the new one and conventional one are combined, $\alpha$ is negative at low $p_{t}$ region, and the photon fragmentation contribution to $\alpha$ becomes more and more important as $p_t$ increase. It makes $\alpha$ go from negative to positive gradually.

In summary, we have suggested a new mechanism for $J/\psi$ production at hadron colliders, $pp(\bar p)\rightarrow \gamma^{\ast}(J/\psi)+X$, with $J/\psi$ from photon fragmentation and calculated the $J/\psi$ hadroproduction from this mechanism up to QCD NLO at the Tevatron and LHC.  Although the obtained total cross section is small, its contribution has large impact on the $p_t$ distribution of $J/\psi$ production and polarization in large $p_{t}$ region. The results show that the $p_t$ distribution for the new production mechanism at NLO is larger than that of the conventional $J/\psi$ production from the CSM at NLO when $p_{t}>26$ (35) GeV at the Tevatron (LHC) and reach about 6 (10) times of the conventional one when $p_t=50$ (100) GeV at the Tevatron (LHC), in spite of a suppression factor $(\alpha/\alpha_s)^2$. In addition, the $p_t$ distribution of $J/\psi$ polarization changes from longitudinal to transverse as $p_t$ increased by adding the new contribution to the conventional one. It is closer to the experimental result than the conventional one. In the future ATLAS and CMS experiments at LHC, the measured $p_t$ distributions of $J/\psi$ production and polarization can be extended to much larger $p_t$ region with its 14TeV center-of-mass energy and high luminosity.  Therefore it is a important mechanism for $J/\psi$ production at large $p_t$ region especially for the LHC.

We would like to thank Dr. Gong Bin for helpful discussion. This work is supported by the National Natural Science Foundation of China (No.10775141) and by the Chinese Academy of Science under Project No. KJCX3-SYW-N2.

\end{document}